\documentclass[a4paper,12pt]{article}
\pdfoutput=1
\usepackage[utf8]{inputenc}
\usepackage[T1]{fontenc}
\usepackage{amssymb}
\usepackage{amsmath}
\usepackage{graphicx}

\DeclareTextSymbol{\degre}{T1}{23}

\begin{document}

\title{\large \bf Entropy production and multiple equilibria: the case of the ice-albedo feedback}

\author{\large C. Herbert$^{ab}$\thanks{Email: corentin.herbert@lsce.ipsl.fr}, D. Paillard$^{a}$ and B. Dubrulle$^{b}$\\
\\
\small $^a$Laboratoire des Sciences du Climat et de l'Environnement,\\
\small IPSL, CEA-CNRS-UVSQ, Gif-sur-Yvette, France.\\
\small $^b$Service de Physique de l'Etat Condens\'e,\\
\small DSM, CEA Saclay, CNRS URA 2464, Gif-sur-Yvette, France. }
\date{\small March 3, 2011}

\maketitle

\begin{abstract}
\small
Nonlinear feedbacks in the Earth System provide mechanisms that can prove very useful in understanding complex dynamics with relatively simple concepts. For example, the temperature and the ice cover of the planet are linked in a positive feedback which gives birth to multiple equilibria for some values of the solar constant: fully ice-covered Earth, ice-free Earth and an intermediate unstable solution. In this study, we show an analogy between a classical dynamical system approach to this problem and a Maximum Entropy Production (MEP) principle view,  and we suggest a glimpse on how to reconcile MEP with the time evolution of a variable. It enables us in particular to resolve the question of the stability of the entropy production maxima. We also compare the surface heat flux obtained with MEP and with the bulk-aerodynamic formula.
\end{abstract}

\section{Introduction}

A very broad class of problems in climate modelling consists of studying the evolution of a particular field (e.g. surface temperature, precipitation,etc) when an external parameter, or \emph{forcing}, is varied. Most of the time, the response to this variation is not linear. Feedbacks can amplify or damp the effect of the initial perturbation.
One of these feedbacks aroused a proficient branch in scientific literature in the 70s', when Budyko and Sellers simultaneously suggested that the interaction between sea ice and climate could have dramatic consequences. Indeed, the higher the global temperature on Earth, the less the ice cover is likely to extend, and thus the lower the albedo. A lower albedo leads in turn to a higher global temperature, and so on and so forth until all the ice is melted. 
Stimulated by this pioneer work, the questions of the stability of the climate as well as the consequences such feedbacks might have for understanding paleoclimates
were extensively studied, using the whole hierarchy of models, from the most simple Energy Balance Models (EBMs) to the complex General Circulation Models (GCMs).

Using 1D EBMs, Budyko and Sellers had found two stable equilibrium positions for the edge of the ice cover, one corresponding to the present climate and one to a fully ice-covered Earth \cite{Budyko1969, Sellers1969}. A large part of the subsequent work was concerned with verifying that these results still held with various different versions of the Budyko-Sellers models, with different heat transport parameterizations, temperature dependance expressions in the planetary albedo, numerical schemes,... \cite[e.g.]{Faegre1972,Schneider1973,Held1974,North1975a,GalChen1976}.
Some elegant analytical solutions were found for these models \cite{Chylek1975,North1975a,North1975b}, and various mathematical methods were applied to determine the stability of the equilibria \cite{Ghil1976,Su1976,Frederiksen1976,Cahalan1979,North1979}.
Owing to the extreme sensitivity of climate to variations in the solar constant found by the first studies, the precise position of the tipping point between present climate and a \emph{deep freezed} Earth was of primary concern. Further investigation by \cite{Lian1977} and \cite{Oerlemans1978} revealed that the sensitivity was much less than initially thought.
A fundamental question raised by these results was that of the \emph{transitivity} of the climate system in Lorenz's terminology \cite{Lorenz1968,Lorenz1970}, and the difference between forced and free fluctuations \cite{Schneider1973,Ghil1976,Fraedrich1978}. For a comprehensive review of the various models, parameterizations and problems pertaining to Energy Balance Models and the ice-albedo feedback, the reader is referred to \cite{North1981}.

In this contribution, we will first give a brief account of the reformulation of these questions with the vocabulary of dynamical system theory: how do multiple equilibria arise from the ice-albedo feedback, what does the bifurcation diagram look like, etc. The model used here is a two box energy balance model with a simplified radiative transfer using the \emph{Net Exchange Formulation} (see e.g. \cite{Dufresne2005}), and a bulk aerodynamic formula for the surface heat flux. In a second step, we draw an analogy between this dynamical system view and the results obtained when predicting the surface heat flux from the Maximum Entropy Production (MEP) principle. The MEP principle, as originally expressed by \cite{Paltridge1975,Paltridge1978,Paltridge1979} for the climate system, provides a variational principle to compute energy fluxes that are not otherwise constrained by the laws of physics. Originally, Paltridge and others applied MEP to the meridional energy transport \cite[e.g.]{Paltridge1975,Paltridge1978,Grassl1981,Gerard1990,Lorenz2001}, but other studies \cite{Ozawa1997,Pujol2002} indicate that it may be valid on the vertical also.

As noticed by \cite{Oerlemans1978},\cite{Crafoord1978} and \cite{Fraedrich1978}, the bifurcation giving birth to multiple equilibria in the case of the ice-albedo feedback has a fundamentally radiative nature, and has nothing to do with transport properties of the atmosphere. This encourages one in thinking that a zero-dimensionnal model is sufficient to capture the structure of the mechanism while avoiding the use of more cumbersome mathematics (namely the Sturm-Liouville theory, required for one-dimensional models such as \cite{Ghil1976}).
Therefore we will restrict ourselves here to this idealized case.
Note also that most of our work could be transposed easily to other feedbacks, like the water-vapour feedback.

\section{The ice-albedo feedback, multiple equilibria and the hysteresis cycle: the dynamical system approach}

\subsection{A simple two-layer EBM using the Net Exchange Formulation}\label{NEFEBMsection}

\begin{figure}
\begin{center}
\includegraphics[width=\textwidth]{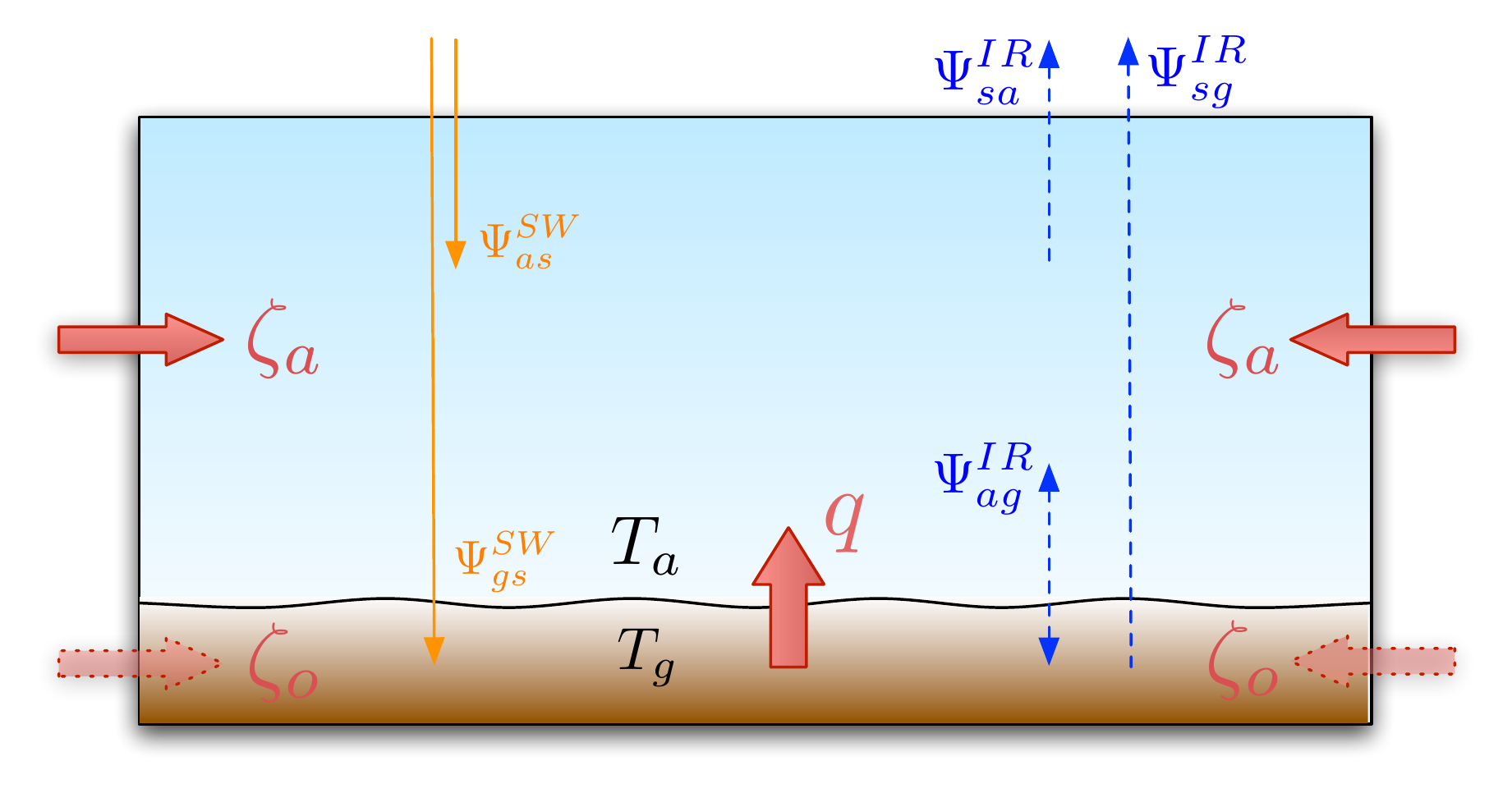}
\caption{A grid cell of the model, adapted from \cite{Herbert2010}. $\Psi_{ij}^\nu$ are the energy exchange rates per unit surface due to radiative transfer (see text), $q$ is the surface heat flux and $\zeta_a$ is the atmospheric energy convergence. Over the oceans, there is also an oceanic energy convergence $\zeta_o$. }\label{model}
\end{center}
\end{figure}

We use a slightly different formulation of the model described in \cite{Herbert2010}, as represented in Fig. \ref{model}. A grid cell is characterized by a surface temperature $T_g$ and an atmospheric temperature $T_a$, and we note $\Psi_{gs}^{SW}$ (respectively $\Psi_{as}^{SW}$) the flux of solar energy received by the ground (respectively absorbed by the atmosphere). 
Radiative exchange use the \emph{Net Exchange Formulation}, in which the basic objects are not energy fluxes at a given level but rather the energy exchange rate between two layers in the atmosphere or between one layer and a boundary surface (see \cite{Dufresne2005}).
$\Psi_{ag}^{IR}$ is the net energy exchange rate between the ground and the atmospheric column per unit surface (ie the greenhouse effect), and $\Psi_{sa}^{IR}$ (respectively $\Psi_{sg}^{IR}$) is the cooling to space term for the atmosphere (respectively the surface).
The net energy exchange rates per unit surface are expressed as functions of $T_g$ and $T_a$ as:
\begin{eqnarray}\label{radeq1}
\Psi_{gs}^{SW}&=&(\bar{s}(\alpha)-s)(1-\alpha)\xi S,\\
\Psi_{as}^{SW}&=&(s+\alpha s^*)\xi S,\label{radeq2}\\
\Psi_{ag}^{IR}&=&t \sigma T_g^4-t\sigma T_a^4,\label{radeq3}\\
\Psi_{sa}^{IR}&=&t \sigma T_a^4,\label{radeq4}\\
\Psi_{sg}^{IR}&=&\left(1-\frac{t}{\mu}\right)\sigma T_g^4,\label{radeq5}
\end{eqnarray}

where $\sigma$ is the Stefan-Boltzmann constant, $\alpha$ is the surface albedo, $t,s,s^*,\bar{s}$ are radiative coefficients, $S$ is the solar constant, $\xi$ accounts for the annual mean zenith angle of the sun and $\mu$ is the Elsasser factor (see \cite{Herbert2010} for a derivation of the equations and a discussion of the coefficients).

In addition to radiation, energy is exchanged due to atmospheric and oceanic transport as well as surface heat fluxes. Let us merge all these energy transfer modes into two variables: $\gamma_a$ (respectively $\gamma_g$) represents the net convergence (the opposite of the divergence) of energy into the atmospheric cell (respectively the surface layer). Writing $\zeta_a$ for the atmospheric convergence (this variable was designated by $\zeta$ in \cite{Herbert2010}), $\zeta_o$ for the oceanic convergence (this was not taken into account in \cite{Herbert2010}), and $q$ for the surface heat flux, we have
\begin{eqnarray}
\gamma_a&=&\zeta_a+q,\\
\gamma_g&=&\zeta_o-q.
\end{eqnarray}

Knowing the convergence of energy in each cell - atmosphere or ground - it is in general not possible without further assumptions to separate the contribution due to surface fluxes, atmospheric transport, and oceanic transport when applicable. Of course, over land, it is reasonable to assume that $\gamma_g$ is just the surface energy flux (ie $\zeta_o=0$) , and then $\gamma_a+\gamma_g$ is the convergence of energy due to the atmospheric flow.
In this study, as we will only use the zero-dimensional version of this model, we will always have $\zeta_a=\zeta_o=0$, and thus $\gamma_a=-\gamma_g=q$.

At steady-state, the energy balance equations for the atmosphere and the surface read
\begin{eqnarray}\label{ebeqn1}
R_a(T_a,T_g)+\gamma_a&=&0,\\
R_g(T_a,T_g)+\gamma_g&=&0\label{ebeqn2},
\end{eqnarray}
where 
\begin{equation}
\begin{split}
R_a(T_a,T_g)=&\Psi_{as}^{SW}+\Psi_{ag}^{IR}-\Psi_{sa}^{IR}\\
=&(s+\alpha s^*)\xi S + t (\sigma T_g^4-2\sigma T_a^4)\\
\end{split}
\end{equation}
and
\begin{equation}
\begin{split}
R_g(T_a,T_g)=&\Psi_{gs}^{SW}-\Psi_{ag}^{IR}-\Psi_{sg}^{IR}\\
=&(\bar{s}-s)(1-\alpha)\xi S -t(\sigma T_g^4-\sigma T_a^4)-\left(1-\frac{t}{\mu}\right)\sigma T_g^4.
\end{split}
\end{equation}

In this form, the steady-state equations (\ref{ebeqn1})-(\ref{ebeqn2}) cannot be solved since $\gamma_a$ and $\gamma_g$ are unkown. In the next section we introduce a parameterization of these quantities as functions of $T_a$ and $T_g$. In section \ref{entropysection}, we use the MEP principle to compute them.

\subsection{The zero-dimensional model with bulk aerodynamic formula}

In the case of a zero-dimensional, two-layer model considered here, the net convergence of energy in the atmospheric box (ie the divergence of the diabatic heating at the surface, $\gamma_a=q=-\gamma_g$) can be simply interpreted as the surface heat flux. In this section, we adopt a bulk aerodynamic formula \cite{PeixotoPOC} to express this flux as a function of the temperatures $T_a$ and $T_g$:

\begin{equation}
\gamma_a=q_{baf}(T_a,T_g)=c_{pa}C_Du_s\left(T_g-T_a\right).
\end{equation}

where $C_D$ is the drag coefficient, $u_s$ is the surface wind speed and $c_{pa}$ is the heat capacity of the atmosphere per unit surface area (similarly $c_{pg}$ is the heat capacity of the ground).
Now the model can be seen as a two-dimensional dynamical system:

\begin{equation}
\left(
\begin{array}{c}
\dot{T_a} \\
\dot{T_g}
\end{array}
\right)=F(T_a,T_g),\label{fads}
\end{equation}

with
\begin{equation}
F(T_a,T_g)=\left(
\begin{array}{c}
F_1(T_a,T_g)\\
F_2(T_a,T_g)
\end{array}
\right),\label{fadsdef}
\end{equation}

and

\begin{eqnarray}
F_1(T_a,T_g)&=&\frac{1}{c_{pa}}(R_a(T_a,T_g)+q_{baf}(T_a,T_g)),\\
F_2(T_a,T_g)&=&\frac{1}{c_{pg}}(R_g(T_a,T_g)-q_{baf}(T_a,T_g)).
\end{eqnarray}

Our main interest here is to find the equilibrium positions of the system, ie the fixed points of the dynamical system, given by the roots of $F$, and to study their stability.
Of course, the dynamics of a two-dimensional dynamical system can be more complex than just a relaxation to an equilibrium position (although it is still rather gentle, see \cite{GHdsbook} for example), contrary to one-dimensional dynamical systems. 
Still, let us note here that the first equation in $F(T_a,T_g)=0$ can be solved algebraically in $T_a$ to obtain a relation $T_a^*=f(T_g^*)$ where $(T_a^*,T_g^*)$ is a fixed point of the system. Thus the number of fixed points of the two-dimensional system is exactly the number of roots of the scalar equation $F_2(f(T_g),T_g)=0$.

For simplicity, we will consider here the projection of the dynamical system (\ref{fads}) onto the $T_g$ axis:
\begin{equation}
\dot{T_g}=F_2(f(T_g),T_g).\label{fadsdefreduced}
\end{equation}

As just explained, this dynamical system, although not mathematically equivalent to the full dynamical system (\ref{fadsdef}), has the same equilibrium positions.
Physically, this simplification is motivated by the fact that the atmosphere can be assumed to reach equilibrium very quickly, hence the evolution of $T_a$ is enslaved by the dynamics of $T_g$. In other words, the system (\ref{fadsdefreduced}) is just the system (\ref{fadsdef}) with $c_{pa}=0$.

\subsection{Multiple equilibria}\label{dseqssection}

\begin{figure}
\begin{center}
\includegraphics[width=\textwidth]{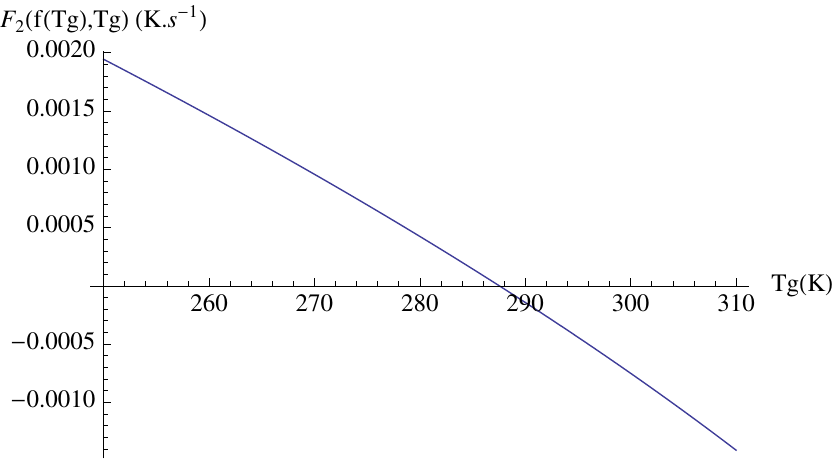}
\caption{Function $F_2(f(T_g),T_g)$ as a function of $T_g$ (see text) with a fixed albedo has only one root.}\label{Frootfixedalb}
\end{center}
\end{figure}

The values of the coefficients used here are reproduced in table \ref{sparamstable}.
Taking for the albedo the fixed value $\alpha_0=0.15$, the system only has one fixed point, as plotting the function $F_2(f(T_g),T_g)=0$ clearly shows (see Fig. \ref{Frootfixedalb}).
In this case, the equilibrium is at a global mean surface temperature of $T_g^0 \approx 288 K$.

\begin{figure}
\begin{center}
\includegraphics[width=\textwidth]{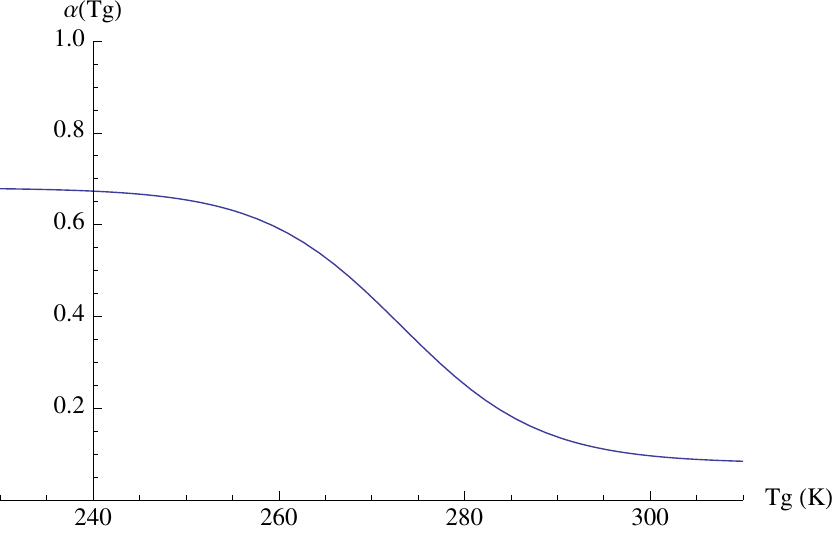}
\caption{Surface albedo $\alpha$ as a function of surface temperature $T_g$ in K.}\label{albgraph}
\end{center}
\end{figure}

But in reality, the higher the global mean temperature, the lower the extent of the regions that can sustain an ice-cover. This positive feedback can be encoded in the following temperature dependance for the albedo :
\begin{equation}
\alpha(T_g)=\alpha_F+\frac{(\alpha_I-\alpha_F)}{2}\left(1+\tanh\left(\frac{T_0-T_g}{\Delta T}\right)\right),\label{iafeq}
\end{equation}

where $\alpha_F$ (respectively $\alpha_I$) represents the value of the planetary albedo over an ice-free (respectively fully ice-covered) area, and $T_0$ and $\Delta T$ are parameters determining the transition from ice-free to ice-covered conditions (see Fig. \ref{albgraph}).
One could simply use a step function between ice-free and ice-covered albedo values, or a piecewise linear function, but we choose this expression because it depends smoothly on the temperature.

Replacing $\alpha$ in Eq. (\ref{fadsdef}) with (\ref{iafeq}) yields a new dynamical system

\begin{equation}
\left(
\begin{array}{c}
\dot{T_a} \\
\dot{T_g}
\end{array}
\right)=G(T_a,T_g),
\end{equation}

where the fixed points are again determined by the conditions, $g$ being defined similarly to $f$ (or obtained by substitution of the albedo function into $f$),

\begin{eqnarray}
T_a^*&=&g(T_g^*), \\
0&=&G_2(g(T_g^*),T_g^*).
\end{eqnarray}

\begin{figure}
\begin{center}
\includegraphics[width=\textwidth]{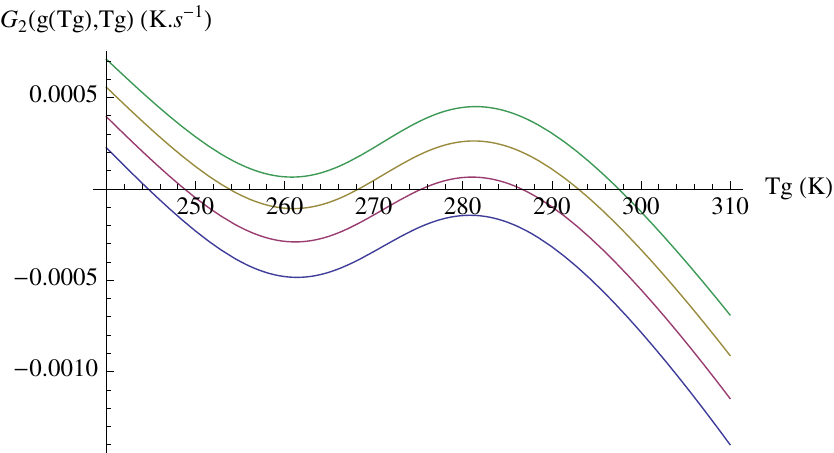}
\caption{Function $G_2(g(T_g),T_g)$ as a function of $T_g$ (see text) including the ice-albedo feedback for different values of the solar constant: $0.95S_0$ (blue), $S_0$ (red), $1.05 S_0$ (yellow), $1.1 S_0$ (green).}\label{Frootsiaf}
\end{center}
\end{figure}

Plotting the curve $G_2(g(T_g),T_g)$ as a function of $T_g$ (Fig. \ref{Frootsiaf}) shows that for certain values of the solar constant, three solutions coexist. This range can be determined to be approximately $0.98 S_0 \leq S \leq1.08 S_0$. Outside this range, only one solution subsists. 
For the present value of the solar constant, $S=S_0$, for instance, these equilibria correspond to a fully glaciated Earth (\emph{snowball} state) $T_g^S \approx 249K$, an ice-free Earth $T_g^P \approx 287K$ which can be identified with the present climate, and an intermediate glacial state $T_g^G \approx 275 K$. For a low value of the solar constant (e.g. $0.95 S_0$), only the snowball state $T_g^S$ subsists. Similarly, at high solar constant (e.g. $1.1 S_0$), the only equilibrium is found on the ice-free branch $T_g^P$.

A fixed point $X^*$ of the dynamical system $\dot{X}=F(X)$ is said to be (linearly) stable if all the eigenvalues of the jacobian of $F$ are negative (see \cite{ArnoldODE} for a complete classification of the two-dimensional fixed points). In this model we find that $T_g^P$ and $T_g^S$ are always stable nodes when they exist, while $T_g^G$ is a saddle-point.

The stability can also be read directly on Fig. \ref{Frootsiaf} for the 1D-reduced system: stable equilibria correspond to roots of the function with negative derivative, while at the unstable equilibrium, the curve crosses the x-axis with an upward slope.

\begin{figure}
\begin{center}
\includegraphics[width=\textwidth]{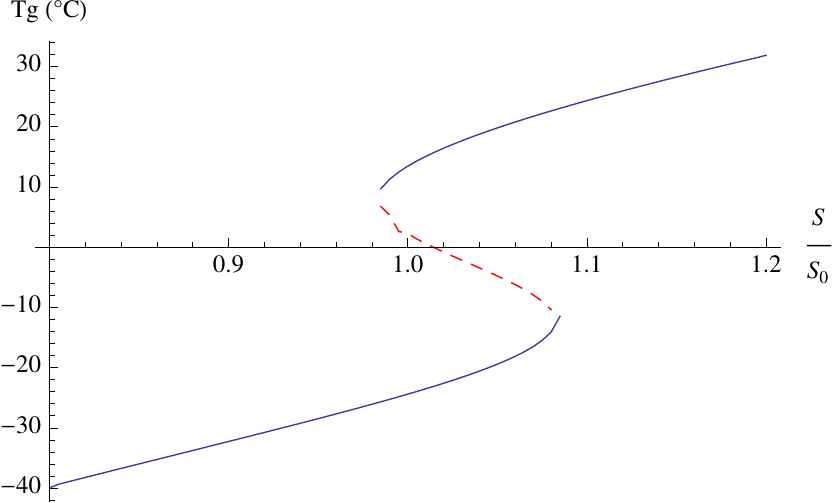}
\caption{Bifurcation diagram of the bulk aerodynamic formula model. $T_G$, $T_P$ and $T_S$ are plotted against $S/S_0$ when they exist. Stable fixed points are plotted in blue while the unstable solution is in dotted red. This figure clearly shows that two saddle-node bifurcations occur at respectively $S\approx 0.98S_0$ and $S \approx 1.08 S_0$. }\label{figbifur}
\end{center}
\end{figure}

Summarizing the above results, Fig. \ref{figbifur} represents the curve of the fixed points when sweeping a large range of values for $S$ : it is the bifurcation diagram of the dynamical system.
Creation of a pair of stable/unstable equilibria at the tipping points $0.98 S_0$ and $1.08 S_0$ is called a \emph{saddle-node bifurcation}. Thus the hysteresis curve obtained for the temperature stems from the bifurcation structure of the dynamical system as two back-to-back saddle-node bifurcations. It is noteworthy that this figure does not depend upon the particular coefficients choice in the bulk formula, nor on the greenhouse effect. Would we set $q_{baf}=0$ (radiative equilibrium with greenhouse effect) or/and $t=0$ (greenhouse effect shut down), the hysteresis curve would remain qualitatively the same.

\subsection{Potential for the dynamical system}\label{dspotentialsection}

The full two-variables dynamical system (\ref{fads}) cannot be expressed as the gradient of a potential function, but its one-dimensional projection can, like any other one-dimensional dynamical system. Let us thus introduce the potential $V$ (defined up to an additive constant) such that

\begin{equation}
\dot{T_g}=-\frac{\partial V }{\partial T_g }.
\end{equation} 

Fixed points of this dynamical system correspond to critical points (ie extrema in this 1D case) of the potential. The stability criterion becomes that stable fixed points are minima of the potential:

\begin{equation}
-\frac{\partial^2 V}{\partial T_g^2 } <0,
\end{equation}

while its maxima are unstable fixed points.

\begin{figure}
\begin{center}
\includegraphics[width=\textwidth]{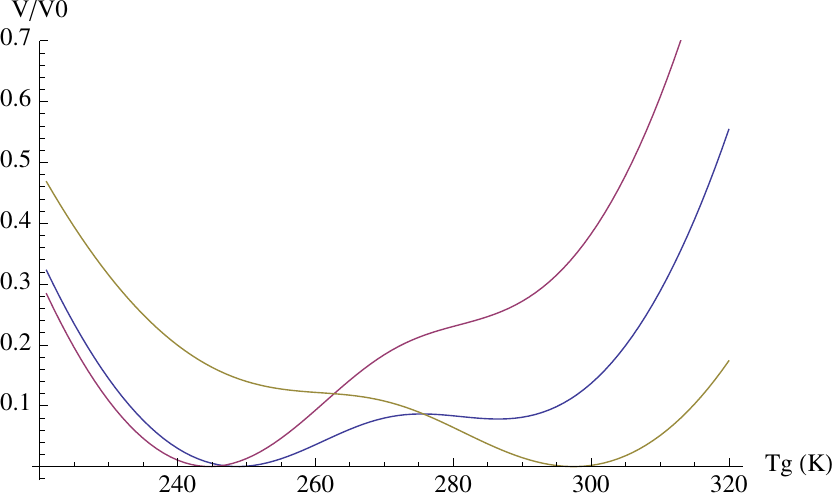}
\caption{Potential $V$ (normalized) as a function of temperature $T_g$ (in K) for three different values of the solar constant: $0.95 S_0$ (red), $S_0$ (blue) and $1.1 S_0$ (yellow). For the present value of the solar constant, the potential has a double well shape, with two stable equilibria, while for the two other values, the potential has only one minimum. }\label{figpotential}
\end{center}
\end{figure}

Figure \ref{figpotential} shows the shape of the potential for different values of the solar constant. At low solar constant (e.g. $0.95 S_0$), the potential has only one critical point, a minimum at $T \approx 245 K$. Increasing the value of the solar constant levels down the potential curve, until a second local minimum appears (along with a local maximum) with $T$ above the freezing point, around $S \approx 0.98 S_0$. At $S=S_0$, it is clear that the potential has two minima at $T\approx250 K$ and $T \approx 290 K$ and a maximum at $T \approx 275 K$. Further increase of the solar constant leads to a deeper minimum at $T>0 \degre$ C while the minimum at $T<0 \degre$ C becomes shallower. Around $S\approx 1.08 S_0$, the minimum at $T<0 \degre$ C disappears (it annihilates with the local maximum) ; for $S=1.1 S_0$, the only minimum is found at $T \approx 300 K $.

Note that, as expected, the critical points of the potential obtained for the three values of the solar constant considered here match with the values of Fig. \ref{Frootsiaf}. Also, the number of critical points of the potential changes at the bifurcation points of the dynamical system.

\section{The entropy production rate and the ice-albedo feedback}\label{entropysection}

In this section, we do not use anymore the bulk aerodynamic formula for the surface flux $\gamma_a=-\gamma_g$, but the Maximum Entropy Production Principle, as described in \cite{Herbert2010}.
The first application of the MEP principle to climate is found in \cite{Paltridge1975}, where the meridional energy transport in a zonally averaged box-model is chosen so as to maximize the entropy production. The resulting climate is in striking accordance with observations. In spite of successful applications in other areas as well, the domain of validity of the MEP principle remains unclear due to the lack of a fully convincing proof (see \cite{Dewar2003,Dewar2005} and the comments in \cite{Grinstein2007,Bruers2007}). More details on theoretical issues and practical use can be found in \cite{Ozawa2003,kleidonlorenzbook,Martyushev2006}.

\subsection{The entropy production rate in zero-dimensions}

Let us consider the model of Sect. \ref{NEFEBMsection} and introduce the entropy production rate per unit surface
\begin{equation}\label{entropyeq}
\sigma= \frac{\gamma_a}{T_{a}}+\frac{\gamma_g}{T_{g}}.
\end{equation}

Substituting Eqs. (\ref{ebeqn1})-(\ref{ebeqn2}) into Eq. (\ref{entropyeq}) for $\gamma_a$ and $\gamma_g$, $\sigma$ can be considered as a functional of the temperature field. We are looking for its maxima subject to the constraint
\begin{equation}\label{constrainteq}
\gamma_a+\gamma_g=0.
\end{equation}

\begin{figure}
\begin{center}
\includegraphics[width=\textwidth]{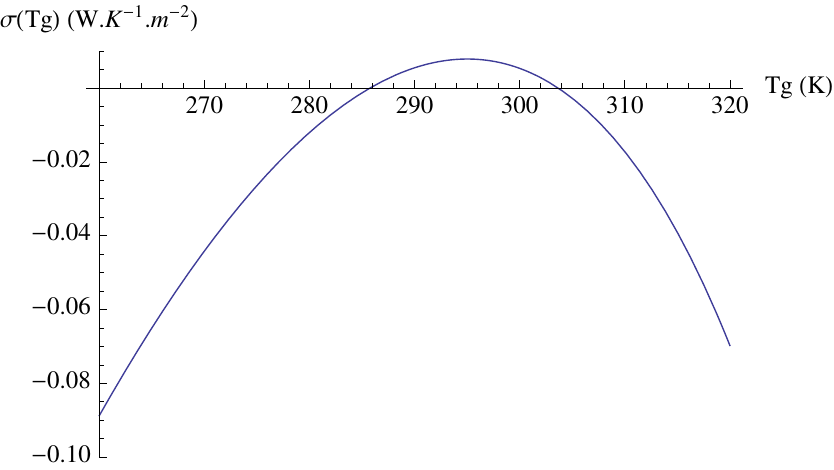}
\caption{Entropy production rate as a function of the surface temperature $T_g$ for the 0D model at $S=S_0$. The only local maximum corresponds to $T_g \approx 295$K. }\label{entropycurve}
\end{center}
\end{figure}

The sum of equations (\ref{ebeqn1}) and (\ref{ebeqn2}) can thus be solved for $T_a$ as a function of $T_g$, and the entropy production rate $\sigma$ is simply a function of one variable.
Its graphical representation for the set of parameters given in table \ref{sparamstable} (fixed albedo $\alpha_0$) is shown in Fig. \ref{entropycurve}. It is clear that there is only one local maximum, corresponding to a surface temperature $T_g \approx 295$K.

\begin{figure}
\begin{center}
\includegraphics[width=\textwidth]{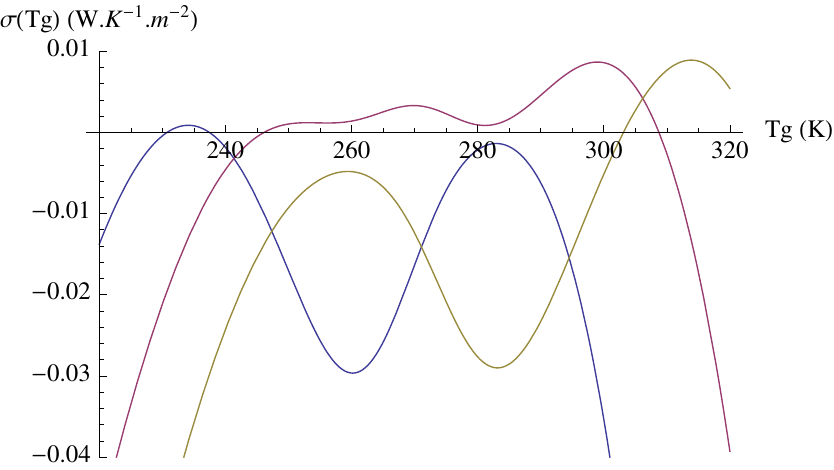}
\caption{Entropy production rate as a function of the surface temperature $T_g$ for the 0D model with ice-albedo feedback. For a low value of the solar constant ($S=0.8 S_0$, blue curve), there is only one local maximum with positive entropy production rate. The same holds for high solar constant ($S = 1.2S_0$, yellow curve), while there are three local maxima and two minima, all with positive entropy production rates, for $S=S_0$ (red curve).}\label{iafentropycurve}
\end{center}
\end{figure}

Now, replacing in the equations the constant albedo $\alpha_0$ by the temperature-dependent albedo (\ref{iafeq}), the resulting entropy production rate curve is plotted in Fig. \ref{iafentropycurve} for different values of the solar constant.

\begin{figure}
\begin{center}
\includegraphics[width=\textwidth]{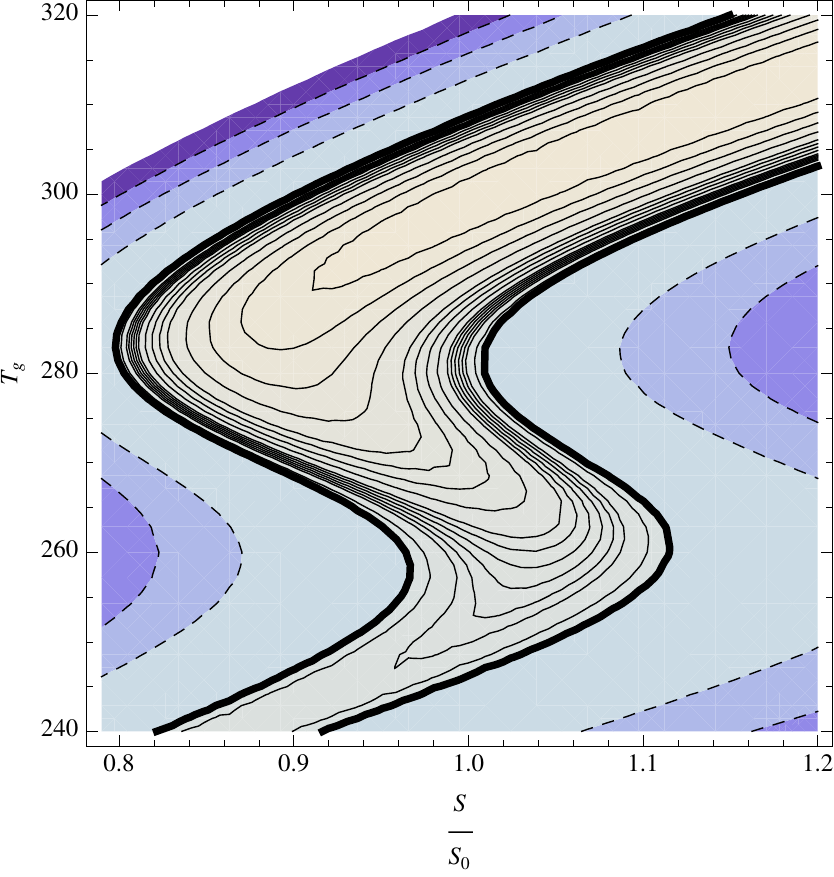}
\caption{Contour plot of the entropy production rate as a function of the solar constant (normalized by its present-day value) and the surface temperature $T_g$ (in K). Negative contour lines are dashed, positive contour lines are solid and the null contour line is the thick solid line. Shades of blue represent negative values of the entropy production rate.}\label{Scontourplot}
\end{center}
\end{figure}

Unlike the potential for the dynamical system in Sect. \ref{dspotentialsection}, the entropy production rate always has at least two local maxima and a local minimum. In fact, over a rather narrow range, estimated to be $0.95 S_0\leq S \leq 1.005 S_0$, the entropy production rate even has three maxima and two minima. This is even clearer on the contour plot of the entropy production rate as a function of $T_g$ and $S/S_0$ (Fig. \ref{Scontourplot}). Hence, there is indeed an analogue of the \emph{fold} of the potential in the classical dynamical system picture in the context of the entropy production surface, but the values at which it takes place do not exactly correspond.

Besides, a large portion of the curve on Fig. \ref{iafentropycurve} lies under the abscissa axis: for the corresponding range of temperature values, the entropy production rate is negative, contrary to what the second law of thermodynamics states (or more precisely its extension to non-equilibrium systems). It seems reasonable to impose the condition 
\begin{equation}
\sigma(T_g) \geq 0,
\end{equation}

thereby restricting the range of values $T_g$ can actually take. In this case, this is equivalent to requiring that the surface heat flux goes from hot to cold.
With this additional constraint, the range of possible values of the solar constant allowing for coexistence of multiple equilibria (two or three) can be determined approximately: $0.8 S_0 \leq S \leq 1.12 S_0$.

\subsection{Stability of the MEP states}\label{mepstabpar}

In the classical understanding of the MEP principle, the rate of entropy production $\sigma$ is a function defined on the manifold of steady-states which reaches a maximum at the most probable state. In the presence of several local maxima, it is generally believed that the final equilibrium state of the system will be the global maximum.
In our case, there are three local maxima of the entropy production rate for the present value of the solar constant, as discussed in the previous section. We know from the dynamical system approach that there can indeed be several steady-states (that coincide with the positions of the entropy production maxima, as discussed in the previous section) for a given set of parameters, and the actual steady-state of the system is determined from the initial conditions. In the absence of fluctuations, the system remains in this state.
Hence it is certainly not sufficient to retain the global maximum of the entropy production rate as representing the final state of the system. Instead one must find a practical way to select a local maximum for given initial conditions. As a particular case, we would obtain a way to distinguish between local entropy production maxima representing \emph{dynamically stable} steady-states and \emph{dynamically unstable} ones.

This involves the introduction of time in the MEP formulation. So far, there was no mention of time in the MEP approach as we were only concerned with steady-states. Even though we claim that the entropy production maxima correspond to equilibrium points, $-\sigma$ is by no means a potential for the dynamical system. Indeed, the dynamics of the system is simply given by the first law of thermodynamics. Here, it reads
\begin{eqnarray}
c_{pa}\frac{dT_a}{dt}&=&R_a(T_a,T_g)+\gamma_a\label{evoleq1},\\
c_{pg}\frac{dT_g}{dt}&=&R_g(T_a,T_g)+\gamma_g\label{evoleq2}.
\end{eqnarray}

Similarly to the steady-state entropy production rate, we can define the instantaneous entropy production rate:
\begin{equation}
\begin{split}
\sigma_i (t)=\frac{\gamma_a (t)}{T_a(t)}+\frac{\gamma_g(t)}{T_g(t)}
=&\frac{1}{T_a(t)}\left(c_{pa}\frac{dT_a}{dt}-R_a(T_a,T_g)\right)\\
&+\frac{1}{T_g(t)}\left(c_{pg}\frac{dT_g}{dt}-R_g(T_a,T_g)\right)
\end{split}
\end{equation}
using Eqs. (\ref{evoleq1})-(\ref{evoleq2}). Note that the instantaneous entropy production rate $\sigma_i$ and the steady-state entropy production rate $\sigma$ coincide at steady-state.

As $\sigma_i$ appears as the natural generalization of $\sigma$ taking into account the time derivative of the dynamical variables, we suggest that the system may follow the trajectory maximizing the instantaneous entropy production rate, seen as a function of the time-dependent unknown fluxes $\gamma_a, \gamma_g$ (always linked by the relation $\gamma_a+\gamma_g=0$). This approach is very similar to what \cite{Jaynes1980} advocates for.

\begin{figure}
\begin{center}
\includegraphics[width=\textwidth]{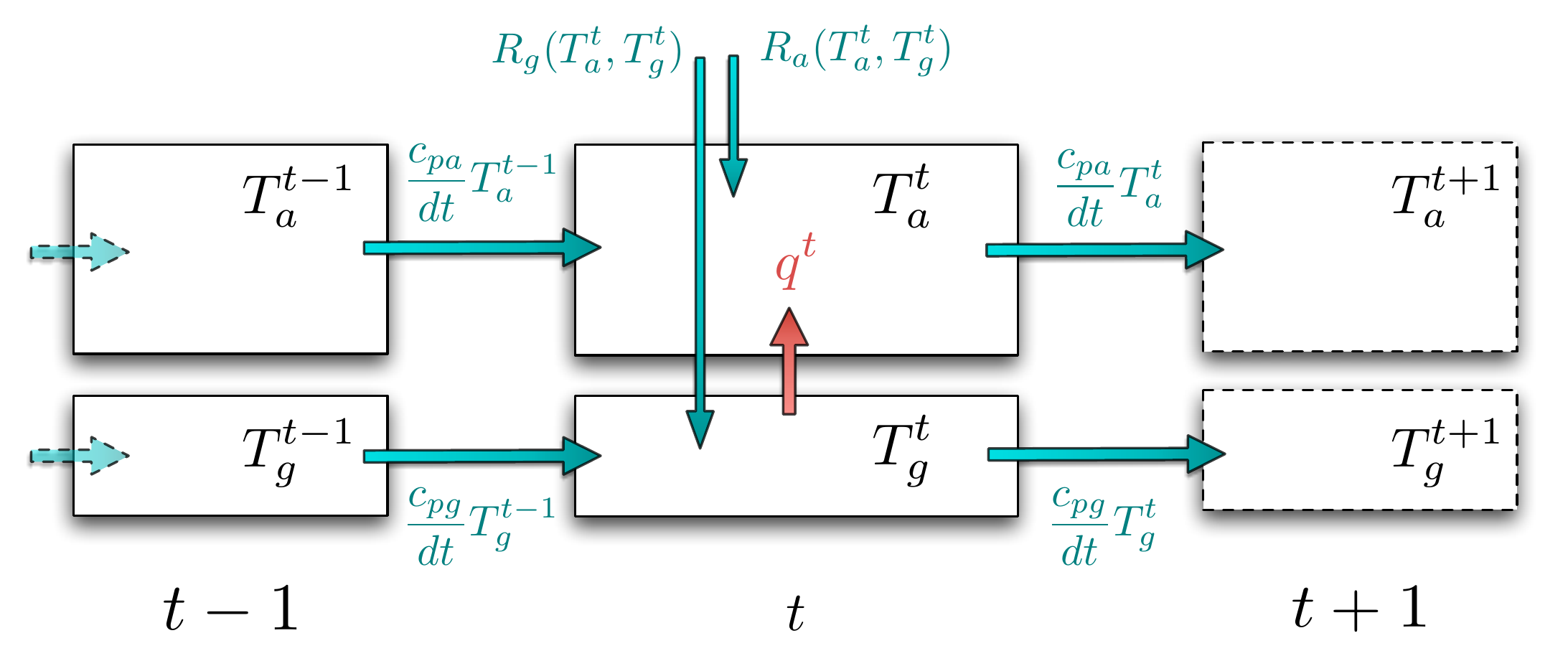}
\caption{To discuss the stability of the steady states predicted by MEP, we need to extend the principle to obtain a time-dependent formulation. This is done by maximizing the instantaneous entropy production rate. To compute the time derivative of the temperature, we consider it as a known flux in time seen as a geometric dimension of the space upon which MEP operates (see text). In green, the fluxes that can be computed from the state variables $(T_a^{t-1},T_a^t,T_g^{t-1},T_g^t)$. In red, the unknown flux obeying MEP.}\label{dynamicMEPfig}
\end{center}
\end{figure}

In practice, it is easier to reformulate the above suggestion with a time-discretized system (see Fig.  \ref{dynamicMEPfig}). Let us consider two snapshots of the system separated by a finite time interval $dt$. We note $T_a^t,T_g^t$ the values of the air and surface temperature at time $t$. The instantaneous entropy production rate becomes:
\begin{equation}
\sigma_i^t=\frac{1}{T_a^t}\left(c_{pa}\frac{T_a^t-T_a^{t-1}}{dt}-R_a(T_a^t,T_g^t)\right)+\frac{1}{T_g^t}\left(c_{pg}\frac{T_g^t-T_g^{t-1}}{dt}-R_g(T_a^t,T_g^t)\right)
\end{equation}
Suppose we know the state of the system at time $t-1$ (ie $T_a^{t-1}$ and $T_g^{t-1}$ are given). Then our postulate is that $T_a^t$ and $T_g^t$ can be chosen so as to maximize $\sigma_i^t$ (with fixed $T_a^{t-1}$ and $T_g^{t-1}$) subject to the constraint $\gamma_a^t+\gamma_g^t=0$. Iterating this process leads to a trajectory maximizing the instantaneous entropy production rate at each timestep, starting from a given initial condition.

\begin{figure}
\begin{center}
\includegraphics[width=\textwidth]{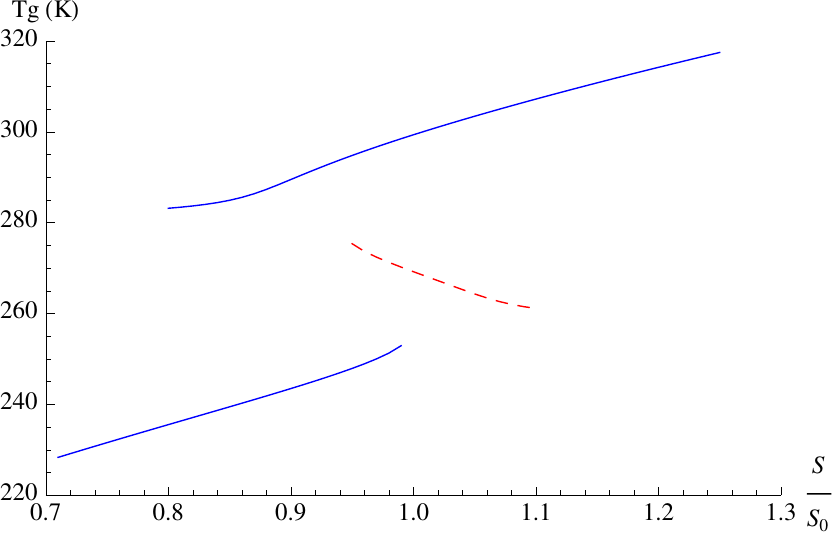}
\caption{Entropy Production maxima as a function of the solar constant, normalized by its present value. The solid lines (respectively the dotted line) correspond to dynamically stable (respectively unstable) equilibria in the sense of paragraph \ref{mepstabpar}. Note that this is not a bifurcation diagram in the usual meaning.}\label{smaxstabfig}
\end{center}
\end{figure}

Integrating the system with this method, initialized in the vicinity of the different maxima of the entropy production rate at steady-state, provides a criterion for stability: it is found here that the warm branch as well as the snowball branch of Fig. \ref{Scontourplot} are stable, while the intermediate branch is unstable. 
The maxima of the entropy production and their stability are plotted as functions of the solar constant on Fig. \ref{smaxstabfig}, analogously to Fig. \ref{figbifur}. This result draws the final line in the parallel between the dynamical system approach and the MEP approach. Note that the limits of this analogy are reached at some points: figure \ref{smaxstabfig} cannot be considered as a usual bifurcation diagram. As a consequence, the lines of existence of the maxima need not depend continuously on the parameter, and for certain values of the parameter (for example $S\approx 0.9 S_0$), two stable maxima coexist with no \emph{unstable manifold} to separate them.

The trajectory maximizing the instantaneous entropy production rate in the way explained above thus yields stability properties for the different steady-states that are consistent with the dynamical system approach. Hence, it seems legitimate to use this hypothesis as a \emph{relaxation equation}, in a similar fashion as \cite{Robert1992}. However, there is no certainty that the system actually follows this maximum instantaneous entropy production trajectory. It would be very valuable to investigate the range of validity of this new application of the MEP principle in future studies, theoretically or on other examples. We can already adduce some material to support our \emph{relaxation equations} approach. In fact, the only novelty as compared to the common use of MEP in the steady-state context is the inclusion of time derivatives of the dynamical variables in the entropy production rate. But one can simply consider these time derivatives as known fluxes, playing exactly the same role as $R_a(T_a,T_g)$ or $R_g(T_a,T_g)$. The only difference is that computing these fluxes requires that we consider a bigger system (here simply the state of the system at times $t-1$ and $t$), even though the number of unknowns in the big system remains the same ($T_a^t$ and $T_g^t$, whereas $T_a^{t-1}$ and $T_g^{t-1}$ are fixed). In this respect, there is no fundamental difference between the time dimension and any geometric dimension, which are customarily included in MEP models.

Alternatively, one could consider the total entropy production rate (ie the integral of the instantaneous entropy production rate over time) as a functional of trajectories and claim that the system follows the trajectory that maximizes this functional subject to the relevant constraints (\cite{Filyukov1967a, Filyukov1967b}, \cite{Filyukov1968} and \cite{MonthusArxiv2010} have developed this idea in the case of Markov chains by maximizing the information entropy as a function of both the probability of the states and that of the transition rates). As we should show in a forthcoming study, this is particularly suitable for periodic phenomena, such as the seasonal cycle. Regarding the stability of the steady-states, we expect this method to yield the same results as the maximum instantaneous entropy production relaxation used here.

\subsection{Surface heat flux and Snowball Earth deglaciation}

In the case of the first section, the surface heat flux is parameterized as a function of $T_a$ and $T_g$. As a consequence of this strong constraint, one could draw a bifurcation diagram for $q_{baf}$ very similar to Fig. \ref{figbifur}, with relatively weak surface heat flux $q_{baf}^S$ for low solar constants (around 20 $W.m^{-2}$), strong surface heat flux $q_{baf}^P$ at high solar constants (around $100 W.m^{-2}$), with an unstable branch $q_{baf}^G$ linking the two.

\begin{figure}
\begin{center}
\includegraphics[width=\textwidth]{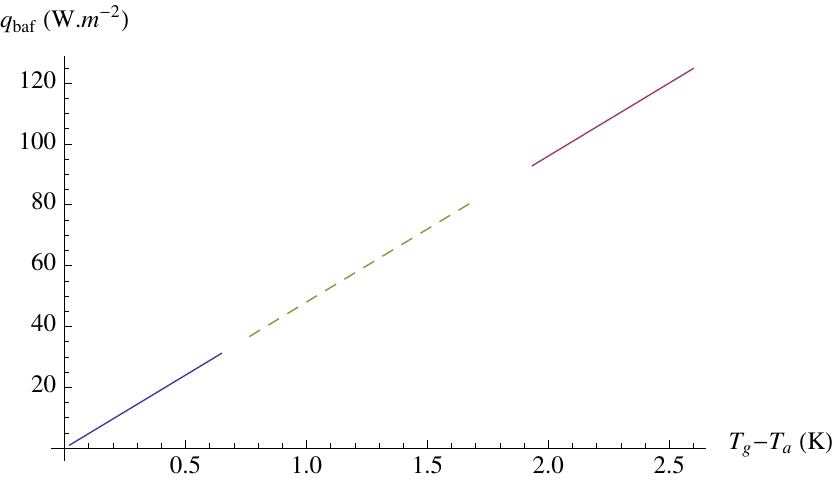}
\includegraphics[width=\textwidth]{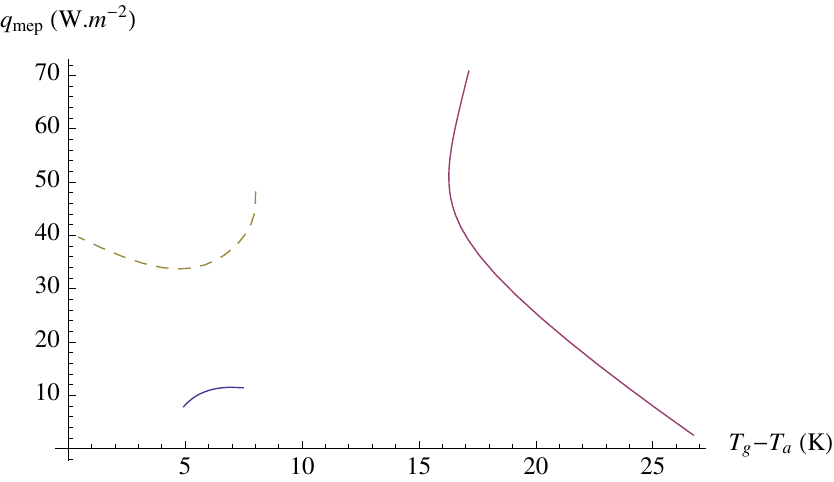}
\caption{Comparison between the bulk aerodynamic formula surface heat flux (top) and the MEP predicted surface heat flux (bottom) as a function of the temperature gradient $T_g-T_a$. The red solid line corresponds to the warm branch of the bifurcation diagram, the blue solid line to the snowball state and the dotted yellow line is the unstable branch. Note the very different scales for $T_g-T_a$.}
\label{qplot}
\end{center}
\end{figure}

On the contrary, the surface heat flux obtained through the MEP procedure $q_{mep}$ is much less constrained by the temperature gradient. Figure \ref{qplot} shows the surface heat flux as a function of the temperature gradient $T_g-T_a$ for both cases: $q_{baf}$ and $q_{mep}$. It is clear that the two differ completely, not only because the temperature gradients in the different climates are very different, but also because the shape of $q_{mep}$ as a function of the temperature gradient is far from linear.
Note that in the MEP snowball state, although the temperature gradient is relatively high, the surface flux remains very low. 
On the warm branch for the MEP state, high values of $q_{mep}$ are obtained for high values of the solar constant. Hence, decreasing the solar constant brings the surface flux down, until the point where only the snowball state survives, with a similar low value of the surface heat flux.

This discrepancy between the two graphs is likely to be significant: it has been suggested that the suppression of the vertical temperature gradient in the snowball state numbers amongst the reasons that make deglaciation of the snowball Earth so difficult \cite{Pierrehumbert2004,Pierrehumbert2005,LeHir2010}. Indeed, the temperature inversion isolates the surface from all the forms of energy exchange: the greenhouse effect can only warm the surface when the air aloft is colder, latent heat plays a very limited role in this very dry atmosphere, and the sensible heat flux is also restricted by the vertical structure of the atmosphere. \cite{Pierrehumbert2004} points out that a crucial role may be played by the surface fluxes parameterization and the convection parameterization.
Here the simplicity of the model does not allow us to discuss the static stability, nor to come up with a clear explanation of the questioning Fig. \ref{qplot}, but it does certainly reinforce the idea that surface heat fluxes parameterization can play critical parts on important paleoclimate problems. In the case of the MEP surface heat flux, our results tend to indicate that it would be possible for the snowball earth to withstand a vertical temperature gradient higher than expected with very little loss in the form of sensible heat, thereby damaging the thermal shield of the surface layer mentioned above. 

On a similar note, \cite{Lucarini2010a} performed a thorough investigation of the thermodynamic properties of the snowball Earth as compared to warm climates in the model of intermediate complexity PLASIM \cite{Fraedrich2005}, using the formalism of non-equilibrium thermodynamics applied to the climate system as described in \cite{Lucarini2009}. Computation of the thermodynamics efficiency, irreversibility and material entropy production clearly characterizes distinct thermodynamic regimes for the snowball Earth and ice-free climate.
Our remarks about the surface heat flux in snowball conditions add up to their thermodynamic analysis.

\section{Conclusion}

The analogy developed in this study leads to some enlightening conclusions. First, about the ice-albedo feedback in itself, it provides a variational principle different from those previously suggested, with a thermodynamic motivation. On the contrary, all the candidates for variational formulations of the problem examined previously were rather \emph{ad hoc} potentials for the dynamical system.
The parallel between potentials properly speaking, which fully describe the dynamics of the system, and the entropy production rate, which only characterize equilibrium states, was pushed one step further with the introduction of a method to integrate a trajectory using the MEP principle. In particular we have shown that this method predicts the correct stability for the MEP predicted equilibria.
We also investigated the behaviour of the surface heat flux in the snowball state. The results hint that MEP might prove useful in such extreme situations where the usual parameterizations face important difficulties. However, the highly simplified model considered here does not allow us to conclude against or in favour of the MEP parameterization, as compared to the bulk-aerodynamic formula.

As far as the MEP conjecture is concerned, our work adds up to the relatively short list of efforts up to now (essentially \cite{Shimokawa2001,Shimokawa2002} and \cite{Jupp2010}) to sort out how the principle should be understood in the presence of multiple entropy production maxima.
\cite{Shimokawa2002} suggested that a dynamical system, in their case the thermohaline circulation, when multiple steady-states are available, should move to the most dissipative one. \cite{Nicolis2003,Nicolis2010} showed strong limitations to this interpretation in full generality. Here, we find that steady-states of a system with \emph{unknown} turbulent fluxes correspond to local maxima of the entropy production seen as a function of the unknown fluxes. The stability of these maxima does not seem to depend on the numeric value of the entropy production at that point. Instead, we suggest that the question of the dynamic stability can be investigated by a relaxation process maximizing the instantaneous entropy production rate.

\bibliographystyle{abbrv}
\bibliography{bibtexlib}

\begin{thebibliography}{10}

\bibitem{ArnoldODE}
V.~Arnold.
\newblock {\em Ordinary Differential Equations}.
\newblock Springer: New-York, 1984.

\bibitem{Bruers2007}
S.~Bruers.
\newblock A discussion on maximum entropy production and information theory.
\newblock {\em Journal of Physics A: Mathematical and Theoretical},
  40:7441--7450, 2007.

\bibitem{Budyko1969}
M.~Budyko.
\newblock The effect of solar radiation variations on the climate of the
  {E}arth.
\newblock {\em Tellus}, 21(5):611--619, 1969.

\bibitem{Cahalan1979}
R.~Cahalan and G.~North.
\newblock A stability theorem for energy-balance climate models.
\newblock {\em J. Atmos. Sci}, 36:1178--1188, 1979.

\bibitem{Chylek1975}
P.~Chylek and J.~Coakley.
\newblock Analytical analysis of a {B}udyko-type climate model.
\newblock {\em J. Atmos. Sci}, 32:675--679, 1975.

\bibitem{Crafoord1978}
C.~Crafoord and E.~K{\"a}ll{\'e}n.
\newblock A note on the condition for existence of more than one steady-state
  solution in {B}udyko-{S}ellers type models.
\newblock {\em J. Atmos. Sci}, 35:1123--1124, 1978.

\bibitem{Dewar2003}
R.~Dewar.
\newblock Information theory explanation of the fluctuation theorem, maximum
  entropy production and self-organized criticality in non-equilibrium
  stationary states.
\newblock {\em Journal of Physics A: Mathematical and General}, 36:631--641,
  2003.

\bibitem{Dewar2005}
R.~Dewar.
\newblock Maximum entropy production and the fluctuation theorem.
\newblock {\em Journal of Physics A: Mathematical and General}, 38:371--381,
  2005.

\bibitem{Dufresne2005}
J.-L. Dufresne, R.~Fournier, C.~Hourdin, and F.~Hourdin.
\newblock Net exchange reformulation of radiative transfer in the $\mbox{CO}_2$
  $15\mu$m band on {M}ars.
\newblock {\em J. Atmos. Sci}, 62:3303--3319, 2005.

\bibitem{Faegre1972}
A.~Faegre.
\newblock An intransitive model of the {E}arth-atmosphere-ocean system.
\newblock {\em J. Appl. Meteorol.}, 11:4--6, 1972.

\bibitem{Filyukov1968}
A.~Filyukov.
\newblock Compatibility property of steady systems.
\newblock {\em Journal of Engineering Physics and Thermophysics},
  14(5):814--819, 1968.

\bibitem{Filyukov1967a}
A.~Filyukov and V.~Karpov.
\newblock Description of steady transport processes by the method of the most
  probable path of evolution.
\newblock {\em Journal of Engineering Physics and Thermophysics},
  13(5):624--630, 1967.

\bibitem{Filyukov1967b}
A.~Filyukov and V.~Karpov.
\newblock Method of the most probable path of evolution in the theory of
  stationary irreversible processes.
\newblock {\em Journal of Engineering Physics and Thermophysics},
  13(6):798--804, 1967.

\bibitem{Fraedrich1978}
K.~Fraedrich.
\newblock Structural and stochastic analysis of a zero-dimensional climate
  system.
\newblock {\em Q. J. R. Meteorol. Soc.}, 104:461--474, 1978.

\bibitem{Fraedrich2005}
K.~Fraedrich, H.~Jansen, E.~Kirk, U.~Luksch, and F.~Lunkeit.
\newblock The planet simulator: Towards a user friendly model.
\newblock {\em Meteorol. Z.}, 14:299--304, 2005.

\bibitem{Frederiksen1976}
J.~Frederiksen.
\newblock Nonlinear albedo-temperature coupling in climate models.
\newblock {\em J. Atmos. Sci}, 33:2267--2272, 1976.

\bibitem{GalChen1976}
T.~Gal-Chen and S.~Schneider.
\newblock Energy balance climate modeling: Comparison of radiative and dynamic
  feedback mechanisms.
\newblock {\em Tellus}, 28(2):108--121, 1976.

\bibitem{Gerard1990}
J.~Gerard, D.~Delcourt, and L.~Francois.
\newblock The maximum entropy production principle in climate models:
  application to the faint young sun paradox.
\newblock {\em Q. J. R. Meteorol. Soc.}, 116:1123--1132, 1990.

\bibitem{Ghil1976}
M.~Ghil.
\newblock Climate stability for a sellers-type model.
\newblock {\em J. Atmos. Sci}, 33(1):3--20, 1976.

\bibitem{Grassl1981}
H.~Grassl.
\newblock The climate at maximum entropy production by meridional atmospheric
  and oceanic heat fluxes.
\newblock {\em Q. J. R. Meteorol. Soc.}, 107:153--166, 1981.

\bibitem{Grinstein2007}
G.~Grinstein and R.~Linsker.
\newblock Comments on a derivation and application of the `maximum entropy
  production' principle.
\newblock {\em Journal of Physics A: Mathematical and Theoretical},
  40:9717--9720, 2007.

\bibitem{GHdsbook}
J.~Guckenheimer and P.~Holmes.
\newblock {\em Nonlinear Oscillations, Dynamical Systems, and Bifurcations of
  Vector Fields}, volume~42 of {\em Applied Mathematical Sciences}.
\newblock Springer: New-York, 1983.

\bibitem{Held1974}
I.~Held and M.~Suarez.
\newblock Simple albedo feedback models of the icecaps.
\newblock {\em Tellus}, 26(6):613--629, 1974.

\bibitem{Herbert2010}
C.~Herbert, D.~Paillard, M.~Kageyama, and B.~Dubrulle.
\newblock Present and last glacial maximum climates as maximum entropy
  production states.
\newblock {\em Q. J. R. Meteorol. Soc., submitted}, 2010.

\bibitem{Jaynes1980}
E.~Jaynes.
\newblock The minimum entropy production principle.
\newblock {\em Annual Review of Physical Chemistry}, 31(1):579--601, 1980.

\bibitem{Jupp2010}
T.~E. Jupp and P.~Cox.
\newblock Mep and planetary climates: insights from a two-box climate model
  containing atmospheric dynamics.
\newblock {\em Philosophical Transactions of the Royal Society B: Biological
  Sciences}, 365:1355--1365, 2010.

\bibitem{kleidonlorenzbook}
A.~Kleidon and R.~Lorenz, editors.
\newblock {\em Non-equilibrium {T}hermodynamics and the {P}roduction of
  {E}ntropy: {L}ife, {E}arth, and {B}eyond}.
\newblock Springer, Berlin, 2005.

\bibitem{LeHir2010}
G.~le~Hir, Y.~Donnadieu, G.~Krinner, and G.~Ramstein.
\newblock Toward the snowball earth deglaciation.
\newblock {\em Climate Dynamics}, 35:285--297, 2010.

\bibitem{Lian1977}
M.~Lian and R.~Cess.
\newblock Energy balance climate models: A reappraisal of ice-albedo feedback.
\newblock {\em J. Atmos. Sci}, 34:1058--1062, 1977.

\bibitem{Lorenz1968}
E.~N. Lorenz.
\newblock Climatic determinism.
\newblock {\em Meteor. Monogr}, 8:1--3, 1968.

\bibitem{Lorenz1970}
E.~N. Lorenz.
\newblock Climatic change as a mathematical problem.
\newblock {\em J. Appl. Meteorol.}, 9(3):325--329, 1970.

\bibitem{Lorenz2001}
R.~Lorenz, J.~Lunine, P.~Withers, and C.~McKay.
\newblock Titan, {M}ars and {E}arth: Entropy production by latitudinal heat
  transport.
\newblock {\em Geophys. Res. Lett}, 28(3):415--418, 2001.

\bibitem{Lucarini2009}
V.~Lucarini.
\newblock Thermodynamic efficiency and entropy production in the climate
  system.
\newblock {\em Phys. Rev. E}, 80:021118, 2009.

\bibitem{Lucarini2010a}
V.~Lucarini, K.~Fraedrich, and F.~Lunkeit.
\newblock Thermodynamic analysis of snowball {E}arth hysteresis experiment:
  Efficiency, entropy production and irreversibility.
\newblock {\em Q. J. R. Meteorol. Soc.}, 136:2--11, 2010.

\bibitem{Martyushev2006}
L.~Martyushev and V.~Seleznev.
\newblock Maximum entropy production principle in physics, chemistry and
  biology.
\newblock {\em Physics Reports}, 426:1--45, 2006.

\bibitem{MonthusArxiv2010}
C.~Monthus.
\newblock Non-equilibrium steady states : maximization of the shannon entropy
  associated to the distribution of dynamical trajectories in the presence of
  constraints.
\newblock arXiv:1011.1342v3, 2010.

\bibitem{Nicolis2003}
C.~Nicolis.
\newblock Comment on the connection between stability and entropy production.
\newblock {\em Q. J. R. Meteorol. Soc.}, 129:3501--3504, 2003.

\bibitem{Nicolis2010}
C.~Nicolis and G.~Nicolis.
\newblock Stability, complexity and the maximum dissipation conjecture.
\newblock {\em Q. J. R. Meteorol. Soc.}, 136:1161--1169, 2010.

\bibitem{North1975a}
G.~North.
\newblock Analytical solution to a simple climate model with diffusive heat
  transport.
\newblock {\em J. Atmos. Sci}, 32(7):1301--1307, 1975.

\bibitem{North1975b}
G.~North.
\newblock Theory of energy-balance climate models.
\newblock {\em J. Atmos. Sci}, 32:2033--2043, 1975.

\bibitem{North1981}
G.~North, R.~Cahalan, and J.~Coakley.
\newblock Energy balance climate models.
\newblock {\em Reviews of Geophysics and Space Physics}, 19:91--121, 1981.

\bibitem{North1979}
G.~North, L.~Howard, D.~Pollard, and B.~Wielicki.
\newblock Variational formulation of budyko-sellers climate models.
\newblock {\em J. Atmos. Sci}, 36:255--259, 1979.

\bibitem{Oerlemans1978}
J.~Oerlemans and H.~van~den Dool.
\newblock Energy balance climate models: Stability experiments with a refined
  albedo and updated coefficients for infrared emission.
\newblock {\em J. Atmos. Sci}, 35(3):371--381, 1978.

\bibitem{Ozawa1997}
H.~Ozawa and A.~Ohmura.
\newblock Thermodynamics of a global-mean state of the atmosphere---a state of
  maximum entropy increase.
\newblock {\em J. Climate}, 10:441--445, 1997.

\bibitem{Ozawa2003}
H.~Ozawa, A.~Ohmura, R.~Lorenz, and T.~Pujol.
\newblock The second law of thermodynamics and the global climate system: A
  review of the maximum entropy production principle.
\newblock {\em Rev. Geophys}, 41(4):1018, 2003.

\bibitem{Paltridge1975}
G.~Paltridge.
\newblock Global dynamics and climate-a system of minimum entropy exchange.
\newblock {\em Q. J. R. Meteorol. Soc.}, 101:475--484, 1975.

\bibitem{Paltridge1978}
G.~Paltridge.
\newblock The steady-state format of global climate.
\newblock {\em Q. J. R. Meteorol. Soc.}, 104:927--945, 1978.

\bibitem{Paltridge1979}
G.~Paltridge.
\newblock Climate and thermodynamic systems of maximum dissipation.
\newblock {\em Nature}, 279:630--631, 1979.

\bibitem{PeixotoPOC}
J.~P. Peixoto and A.~H. Oort.
\newblock {\em Physics of {C}limate}.
\newblock Springer: New-York, 1992.

\bibitem{Pierrehumbert2004}
R.~Pierrehumbert.
\newblock High levels of atmospheric carbon dioxide necessary for the
  termination of global glaciation.
\newblock {\em Nature}, 429:646--649, 2004.

\bibitem{Pierrehumbert2005}
R.~Pierrehumbert.
\newblock Climate dynamics of a hard snowball earth.
\newblock {\em J. Geophys. Res}, 110:D01111, 2005.

\bibitem{Pujol2002}
T.~Pujol and J.~Fort.
\newblock States of maximum entropy production in a one-dimensional vertical
  model with convective adjustment.
\newblock {\em Tellus}, 54:363--369, 2002.

\bibitem{Robert1992}
R.~Robert and J.~Sommeria.
\newblock Relaxation towards a statistical equilibrium state in two-dimensional
  perfect fluid dynamics.
\newblock {\em Physical Review Letters}, 69(19):2776--2779, 1992.

\bibitem{Schneider1973}
S.~Schneider and T.~Gal-Chen.
\newblock Numerical experiments in climate stability.
\newblock {\em J. Geophys. Res}, 78(27):6182--6194, 1973.

\bibitem{Sellers1969}
W.~Sellers.
\newblock A global climatic model based on the energy balance of the
  earth-atmosphere system.
\newblock {\em J. Appl. Meteorol.}, 8:392--400, 1969.

\bibitem{Shimokawa2001}
S.~Shimokawa and H.~Ozawa.
\newblock On the thermodynamics of the oceanic general circulation: entropy
  increase rate of an open dissipative system and its surroundings.
\newblock {\em Tellus A}, 53(2):266--277, 2001.

\bibitem{Shimokawa2002}
S.~Shimokawa and H.~Ozawa.
\newblock On the thermodynamics of the oceanic general circulation:
  Irreversible transition to a state with higher rate of entropy production.
\newblock {\em Q. J. R. Meteorol. Soc.}, 128:2115--2128, 2002.

\bibitem{Su1976}
C.~Su and D.~Hsieh.
\newblock Stability of the budyko climate model.
\newblock {\em J. Atmos. Sci}, 33(12):2273--2275, 1976.

\end{thebibliography}

\clearpage

\begin{table}[t]
\caption{Values for the parameters of the 0D model (radiative coefficients, bulk aerodynamic formula parameters and ice-albedo feedback parameterization). Note that the values for the heat capacities depend on the thickness of the layer and on the nature of the surface (ocean or land), but this has no influence on the steady-state results.}\label{sparamstable}
\vskip4mm

\begin{tabular}{c*{7}{c}}
\hline
 $\mu$ & $\xi$ & $t$ & $s$ & $s^*$ & $\bar{s}$ &$\alpha_0$& $S_0$\\
 0.6 & 0.25 & 0.44 & 0.19 & 0.015 & 0.89  & 0.15 & 1368 $W.m^{-2}$\\
\hline
 $C_D$ & $u_s$ & $c_{pa}$ & $c_{pg}$ & $\alpha_I$ & $\alpha_F$ & $T_0$ & $\Delta T$ \\
 0.008 & 6 $m.s^{-1}$ & 1 $MJ.K^{-1}.m^{-2}$ & 210. $MJ.K^{-1}.m^{-2}$  & 0.08 & 0.68 &273.15K& 15K\\
\hline
\end{tabular}
\end{table}

\end{document}